# Hybrid Photonic-Plasmonic Cavities based on the Nanoparticle-on-a-Mirror Configuration

Angela I. Barreda,[†,‡] Mario Zapata-Herrera,[fj,§] Isabelle Palstra,[‖,⊥] Laura Mercadé,[‡] Javier Aizpurua,[fj,§] A. Femius Koenderink,[‖,⊥] and Alejandro Martínez[*,‡]

[†]*Institute of Applied Physics, Abbe Center of Photonics, Friedrich Schiller University Jena, Albert-Einstein-Str. 15, 07745 Jena, Germany*
[‡]*Nanophotonics Technology Center, Universitat Politècnica de València, Camino de Vera s/n, 46022, Valencia, Spain*
[fj]*Materials Physics Center CSIC-UPV/ EHU, 20018 Donostia-San Sebastian, Spain*
[§]*Donostia International Physics Center DIPC, 20018 Donostia-San Sebastian, Spain*
[‖]*Center for Nanophotonics, AMOLF, Science Park 104, 1098 XG Amsterdam, The Netherlands*
[⊥]*Van der Waals-Zeeman Institute, Institute of Physics, University of Amsterdam, Science Park 904, PO Box 94485, 1090 GL Amsterdam, The Netherlands*

E-mail: amartinez@ntc.upv.es

**Abstract**

Hybrid photonic-plasmonic cavities have emerged as a new platform to increase light-matter interaction capable to enhance the Purcell factor in a singular way not attainable with either photonic or plasmonic cavities separately. In the hybrid cavities proposed so far, mainly consisting of metallic bow-tie antennas, the plasmonic gap sizes defined by lithography in a repeatable way are limited to minimum values ≈ 10 nm. Nanoparticle-on-a-mirror (NPoM) cavities are far superior to achieve the smallest possible mode volumes, as gaps smaller than 1 nm can be created. Here, we design a hybrid cavity that combines a NPoM plasmonic cavity and a dielectric-nanobeam photonic crystal cavity operating at transverse-magnetic (TM) polarization. The metallic nanoparticle can be placed very close (< 1 nm) to the upper surface of the dielectric cavity, which acts as a low-reflectivity mirror. We demonstrate through numerical calculations that this kind of hybrid plasmonic-photonic cavity architecture exhibits quality factors, $Q$, above $10^3$ and normalized mode volumes, $V$, down to $10^{-3}$, thus resulting in high Purcell factors ($F_P \approx 10^5$), whilst being experimentally feasible with current technology. Our results suggest that hybrid cavities with sub-nm gaps should open new avenues for boosting light-matter interaction in nanophotonic systems.

## Introduction

Optical cavities are structures tailored to localize light in small volumes during long residence times. To quantify light-matter interaction in optical cavities, a key metric is the Purcell factor ($F_P$),[1–3] which is proportional to the ratio between the quality factor $Q$ (or $Q$-factor) and the effective volume of the cavity mode, quantifying the local density of optical states (LDOS) at resonance

$$F_P = \frac{3}{4\pi^2} \frac{Q}{V} \qquad (1)$$

where $V$ is mode volume normalized by the wavelength over the local refractive index cubed, $(\lambda/n)^3$ and the $Q$-factor measures the lifetime of a photon in the cavity in units of optical cycles. Dielectric cavities made of transparent materials enable very large $Q$ values[4] since photons can remain for a long time inside the cavity without dissipation.[5] However, the mode volume, which accounts for the spatial extension of the electromagnetic field inside a cavity, has a diffraction-limited floor $V \approx 1$. Thus, reaching subwavelength confinement ($V < 1$) requires the use of metallic cavities with nm-scale plasmonic gaps.[6,7] This allows to overcome the diffraction limit at the expense of having low $Q$-factors (around 10) as a result of the large absorption and scattering losses provided by the metal. Basically, this constrains optimization of the Purcell factor: when increasing cavity storage times in a dielectric cavity, the spatial confinement cannot be extreme. Conversely, extreme subwavelength spatial confinement in nm-scale plasmonic gaps comes at the cost of very short photon lifetimes, thus reducing the $Q$-factor.[8,9]

In the last few years, hybrid plasmonic-photonic cavities[8,10–16] have emerged as a promising way of mixing both types of confinement approaches, taking advantage of the idea of placing a gap plasmonic nanoantenna in a large field confinement region of a dielectric cavity where both modes, plasmonic and photonic, can hybridize. This results in new features not attainable by either plasmonic or photonic cavities when operating individually. Interestingly, such cavities enable tuning both the $Q$-factor and the normalized mode volume $V$ by properly engineering the coupling between the plasmonic and photonic modes.[11] This also allows to harness the value of the Purcell factor so that it can be suitably chosen for different applications in classical and quantum optics. For the latter, previous works have mainly focused on single-photon sources, low-threshold lasers, strong coupling with quantum emitters or sensing and vibrational spectroscopy.[11,17–19]

Among the different possible implementations of such hybrids, the integration of bow-tie nanoantennas as canonical plasmonic structures on dielectric photonic crystal cavities has a set of advantages,[11] such as the low value of the mode volume of the photonic cavity and the possibility of fabrication using conventional lithography methods. Noticeably, such hybrid cavities have not yet witnessed an experimental demonstration, although several experiments have confirmed the integration of bow-tie nanoantennas in dielectric waveguides.[20–25] A main limitation of this hybrid plasmonic-photonic cavity approach is that the bow-tie gap is defined by lithography, which limits its minimum attainable value. Whilst reaching gap widths below 10 nm is attainable,[26–28] repeatability is very poor: given a same nominal value, the obtained gap width is extremely dependent on the local conditions and can take different values - or get closed - for different bow-tie nanoantennas, even when fabricated in same lithography step.[21,28] In general, reaching nm- and sub-nm plasmonic gaps in a controllable fashion becomes extremely complex and hardly reconcilable with the multistep lithography challenge of integration with photonic cavities.

A much more appealing approach to reach nm- and sub-nm-scale plasmonic gaps in a repeatable way is by vertical deposition. Within this paradigm, nanoparticle-on-a-mirror (NPoM) plasmonic cavities have demonstrated unrivalled performance in extreme spatial field confinement.[29–31] The smallest demonstration of mode volume so far has been in picocavities, where the electromagnetic field is ultimately confined around a single metallic atom.[7,32,33] By using molecular monolayers, or atomically thin layers, so-called nanocavities can be routinely achieved, i.e. mode volumes $\leq \lambda^3/10^5$.

In order to hybridize such gap modes as exist in vertically assembled NPoM structures with a photonic crystal cavity, one would require a confined photonic mode with the main component of the electric field pointing along the vertical direction. Unfortunately, photonic crystal cavities usually operate with transverse-electric (TE) modes, which have the main component of the transversal electric field pointing in-plane. This is because TE bandgaps are more easily obtained than their transverse magnetic (TM) counterparts when implemented by drilling holes in thin ($\approx$ 200 nm) high-index semiconductor films.[34] In this work, we introduce a novel class of hybrid cavity resulting from the hybridization between a metallic nanoparticle and a photonic crystal cavity that supports a highly-confined TM mode at $\lambda_c \approx$ 700 nm. We show that the upper surface of the photonic crystal can act as a low-reflectivity mirror when the metallic nanoparticle is placed on top of it. For nanoparticle-mirror gaps ($d$) around 1 nm, we observe a strong reduction of the mode volume without a significant impairment of the $Q$-factor of the photonic cavity. Compared to the Purcell factors acquired with feasible gaps ($d \approx$ 10 nm) for hybrid plasmonic-photonic configurations obtained from TE photonic modes, we reach an improvement of one order of magnitude. We also discuss the possible experimental design, which would require the use of materials with high refractive index ($n >$ 3) and negligible losses at visible and near-infrared spectral ranges, being gallium phosphide a feasible candidate.

## Description of the hybrid system

The system under study in this work is schematically depicted in the top panel of Fig. 1: A metal nanoparticle is placed on top of a one-dimensional photonic crystal cavity created in a high-index dielectric nanobeam. These two elements are depicted separately in the left and right panels of Fig. 1. The goal of the hybridization is to improve the quality factor of the plasmonic nanoparticle by suppressing scattering pathways as a result of the photonic bandgap of the dielectric cavity, and to simultaneously decrease the mode volume of the dielectric cavity by squeezing the field with the metal nanoparticle. Note that, unlike in the standard NPoM configuration where the mirror is a metallic surface, in the proposed hybrid structure, it is the upper interface of the photonic crystal what acts as a mirror. This means that a high-index of refraction is required to increase the reflectivity of the interface and boost the light confinement in the spacing gap. Moreover, in the current scheme, extreme light confinement will take place for the electric field pointing from the upper interface towards the metal nanoparticle. This means that the dielectric cavity must operate for TM polarization (or odd-parity modes in the context of photonic crystal slabs[35,36]). Obtaining odd-parity photonic bandgaps is not easy in thin semiconductor nanobeams drilled by holes, which usually tend to have bandgaps for TE modes.[35] However, by using thick substrates made of high-index dielectric materials (such as silicon), the realization of high-Q cavities for odd (or TM) modes by drilling nanoholes becomes feasible.[37,38] Nevertheless, in order to operate at visible or near-infrared wavelengths, where many applications in the contexts of cavity quantum electrodynamics can be found, silicon is not allowed and there are not so many transparent materials with a large index of refraction. Recently, gallium phosphide has been suggested as an interesting optical material for nonlinear[39] or optomechanical[40] applications. Interestingly, it shows a large refractive index ($n > 3$) and is transparent at wavelengths over $\lambda$ = 570 nm. For those reasons, this is the material that we have chosen to implement the photonic crystal cavity. In order to evaluate the effects of the dielectric substrate on the optical properties of the metal nanoparticle (when place in its close proximity), we also consider the intermediate system depicted in the bottom panel of Fig. 1. The analysis of this structure will allow us to separate the effects induced by the non-structured dielectric substrate, merely acting as a "bad" mirror, and the whole perforated cavity having a bandgap for TM-polarized light.

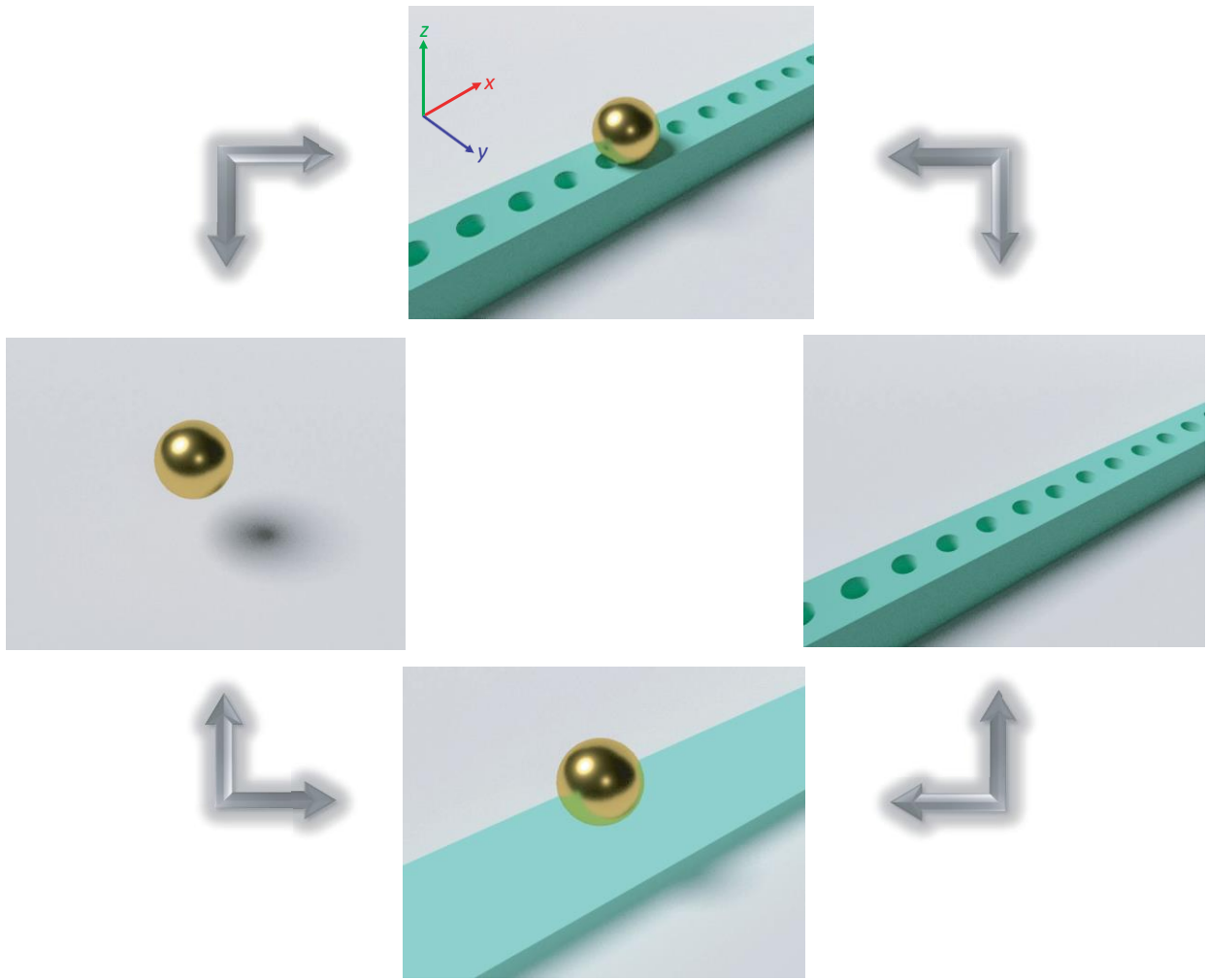

Figure 1: Conceptual scheme of the hybrid system under study. (Top) A metallic nanoparticle is placed on top of a dielectric cavity. Both structures are spaced by a tiny gap of thickness *d*. The isolated systems are a metallic nanoparticle (left), in this case a gold nanosphere; and a photonic crystal cavity (right) having a photonic bandgap for TM modes. An intermediate system (bottom) arises when the metal nanoparticle is placed on top of a non-structured dielectric medium.

# Numerical Results

We start by considering the electromagnetic response of an isolated metallic nanosphere, which for small radius exhibits a dipolar plasmonic response at visible wavelengths. In Fig. 2a, we show the calculated normalized local optical density of states (LDOS) for a gold nanosphere of radius $R$ = 40 nm (see Methods for details about the LDOS calculations), which displays a dipolar electric resonance at $\lambda_{NP} \approx 500$ nm. For all the results shown in this manuscript, the LDOS is calculated by considering an electric dipole (whose dipole moment is oriented along the $z$-axis, i.e., along the line connecting the sphere and the cavity) centered at 0.5 nm below the sphere and 0.5 nm above the surface of the beam as the illumination source. In analogy to Ref. 11, LDOS values are normalized to the LDOS in vacuum at the same wavelength, with LDOS containing the radiative and nonradiative contributions. The optical quality factor $Q_{NP}$ = 17.45 (retrieved by means of a Fano lineshape fitting) and the normalized mode volume $V_{NP}$ = 2.2 × 10$^{-6}$ are shown in the inset. The normalized mode volume $V$ is obtained by using the value of the normalized LDOS at the peak of the resonance, and then applying Eq. 1. Noticeably, $V$ becomes very small, as expected from plasmonic cavities, far smaller than the diffraction limit. The extreme confinement is clearly appreciated in Fig. 2b, which depicts the near-field map corresponding to the $z$-component of the electric field under resonance, with linearly polarized light (along the $z$-axis) propagating along the $x$-axis. This figure corroborates the dipolar character of the excited mode at resonance in Fig. 2a.

When the photonic crystal cavity is considered, the $Q$ vs. $V$ situation is reversed. Our dielectric cavity consists of a gallium phosphide (GaP with refractive index $n$ = 3.49) nanobeam of width $w$ = 200 nm and thickness $t$ = 250 nm drilled with circular holes. To build the cavity, we calculate the parameters to get a TM band-gap around $\lambda \approx 700$ nm, which happens for a period $P_M$ = 165 nm and hole radii of $R_M$ = 43.725 nm. We choose this wavelength because, as shown in previous works,[11] the nanoparticle resonance must be blue-shifted regarding that of the photonic cavity to improve the performance of the hybrid cavity with respect to the bare components. Nevertheless, the resonance wavelength could be either red- or blue-shifting by changing the parameters defining the photonic crystals. With the previous dimensions, we form two TM mirrors with ten holes each at every side of the cavity. The photonic cavity is subsequently formed by adiabatically changing the dimensions of the mirrors when moving towards the cavity center, where the nanoparticle will be placed. In particular, we reduce the period and hole size of the photonic crystal during seven holes at each side of the cavity by means of an quadratic adiabatic transition down to $P_D$ = 138.6 nm and $R_D$ = 36.729 nm, respectively.[37,38] This leads to a confined TM mode with a large Q factor, as shown in Fig. 2c, which depicts the obtained normalized LDOS for the isolated photonic cavity, calculated by considering the illumination of an electric dipole placed 0.5 nm above the cavity. From the fitting of the data to a Lorentzian lineshape, the $Q$-factor and normalized mode volume $V$ values are extracted. In this case we obtain a large $Q$-factor ($Q_C$ = 1.54 × 10$^4$), which could be further engineered to values even above 10$^5$, and a diffraction-limited normalized volume $V_C \approx 1$. Looking at the mode profile of the $z$-component of the electric field distribution in the cavity at resonance (see Fig. 2d) (ZX crosscut of the cavity design corresponding to the length-thickness plane at the middle of the beam width), we can observe that it has evanescent tails of $E_z$ on top of the upper interface, which may couple to the evanescent field perpendicular to the nanosphere surface (Fig. 2b) thus giving rise to the pursued hybridization of the plasmonic and photonic resonances. The field distribution, calculated by using the COMSOL Multiphysics eigenmode solver, corresponds to the fundamental mode (see Methods section for further details). Henceforth all the near-field maps shown in this work are obtained using this methodology.

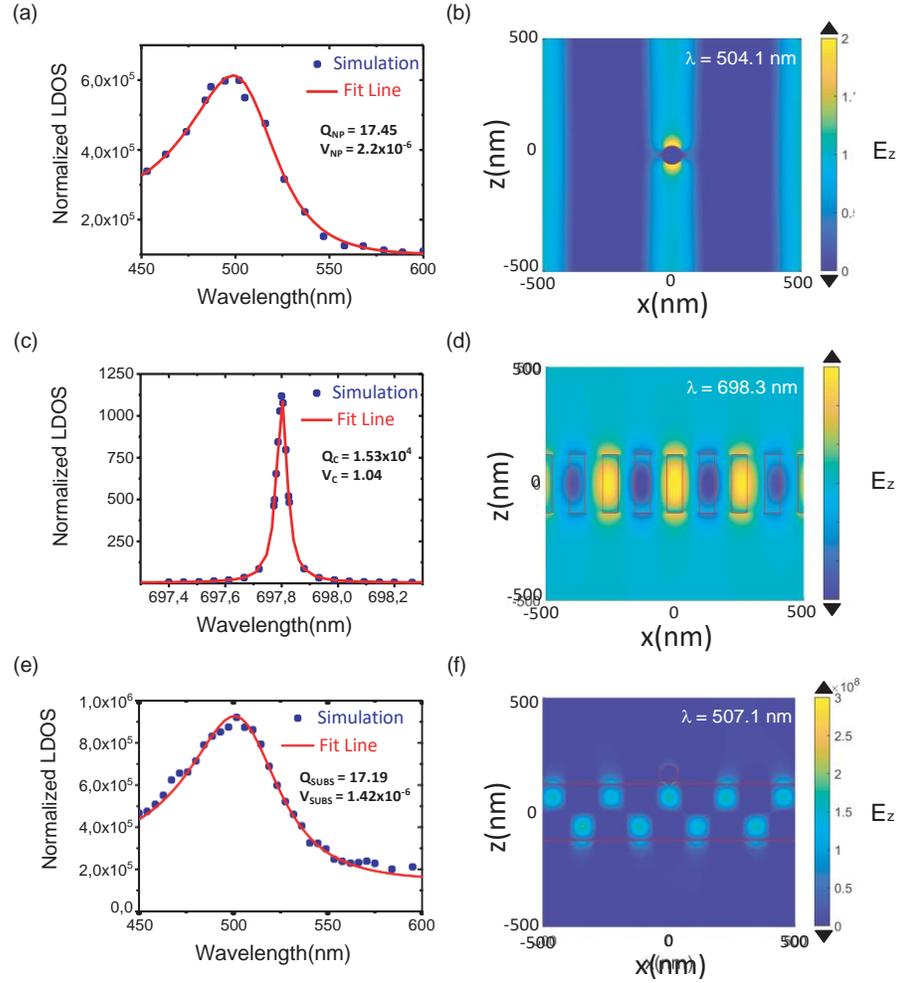

Figure 2: Simulation results of the different building blocks of the hybrid cavity. (a) Normalized LDOS, quality factor and normalized mode volume and (b) Mode profile (ZX crosscut) of the *z*-component of the electric field for a gold nanoparticle of radius $R$ = 40 nm. (c,d) are as (a,b) but for the dielectric gallium phosphide photonic crystal cavity described in the main text. (e,f) are as (a,b) but assuming that the nanosphere ($R$ = 40 nm) is placed on top of a non-structured gallium phosphide dielectric slab (width $w$ = 200 nm and thickness $t$ = 250 nm).

In order to estimate the effect of the dielectric layer acting as a low-reflectivity mirror, we also simulate the response of the metallic nanosphere when located on the top of an non-structured GaP nanobeam (width $w$ = 200 nm and thickness $t$ = 250 nm) to determine the influence of the dielectric substrate on the electromagnetic behavior of the gold nanoparticles. In other words, we simulated the same geometry as in the case of the hybrid cavity (below) but without holes. The gap between the metallic sphere and the dielectric slab corresponds is $d$ = 1 nm, a value that is kept in all our calculations. For this configuration, the $Q$-factor ($Q_{SUBS}$ = 17.19) and normalized mode volume $V$ ($V_{SUBS}$ = 1.42 × $10^{-6}$) are similar to those obtained for the isolated sphere. This evidences that the GaP substrate behaves as a "bad" mirror as well as the necessity of including the photonic crystal to build up the hybrid system.

To form the hybrid cavity, we place the gold nanosphere on top of the photonic crystal beam (top panel, Fig. 1) keeping $d$ = 1 nm. In real experiments, this spacer would be filled by a self-assembled monolayer (SAM). To understand how the detuning between the cavity and nanosphere resonant wavelengths affects the results, different radii for the metallic nanosphere are considered, varying from $R$ = 30 nm to 70 nm in 10 nm steps. In Fig. 3a and b, the normalized LDOS spectra, the $Q$-factor and the normalized mode volume values $V$ for the different analyzed hybrid systems are shown. Again, the LDOS is obtained by considering the illumination of an electric dipole placed 0.5 nm above the cavity, whose dipole moment is oriented along the $z$-axis. The parameters of the hybrid, $Q_{HYB}$ and $V_{HYB}$, are retrieved through the fitting of the LDOS to a Fano lineshape. By comparing the results for the hybrid and the bare systems, it is observed how the hybrid $Q_{HYB}$ and $V_{HYB}$ take intermediate values between those of the photonic cavity and the nanoparticle. This means that, as expected, the hybrid cavity enables small volumes (order $10^{-3}$) below the diffraction limit whereas $Q_{HYB}$ remains relatively high (order $10^3$). This is an indicator of the hybridization of the cavity and nanoparticle responses. The hybridization effect is clearly observed in Fig. 3c, where we represent the mode profile (ZX crosscut) of the $z$-component of the electric field for a nanosphere-based hybrid cavity with $R$ = 40 nm. The extreme concentration of the electric field in the gap, mimicking what takes place in standard NPoM systems, is shown in the inset. In Fig.3a we also observe that as the gold nanoparticle radius increases, its resonant wavelength is red-shifted and the detuning between the metallic nanosphere and the nanobeam cavity decreases. As a result, the resonances are broadened and both the $Q$-factor and normalized mode volume decrease. In fact, the results for $R$ = 30 nm corresponds to $Q_{HYB} = 1.9 \times 10^4$ and $V_{HYB} = 5.1 \times 10^{-3}$ whilst for $R = 70$ nm, they are $Q_{HYB} = 1.6 \times 10^3$ and $V_{HYB} = 1.1 \times 10^{-3}$. We also performed simulations for smaller spheres (not shown). In particular, for $R = 20$ nm we get that $V_{HYB} \approx 10^{-2}$, which means that strong subwavelength spatial confinement arisen from the plasmonic response is eventually lost.

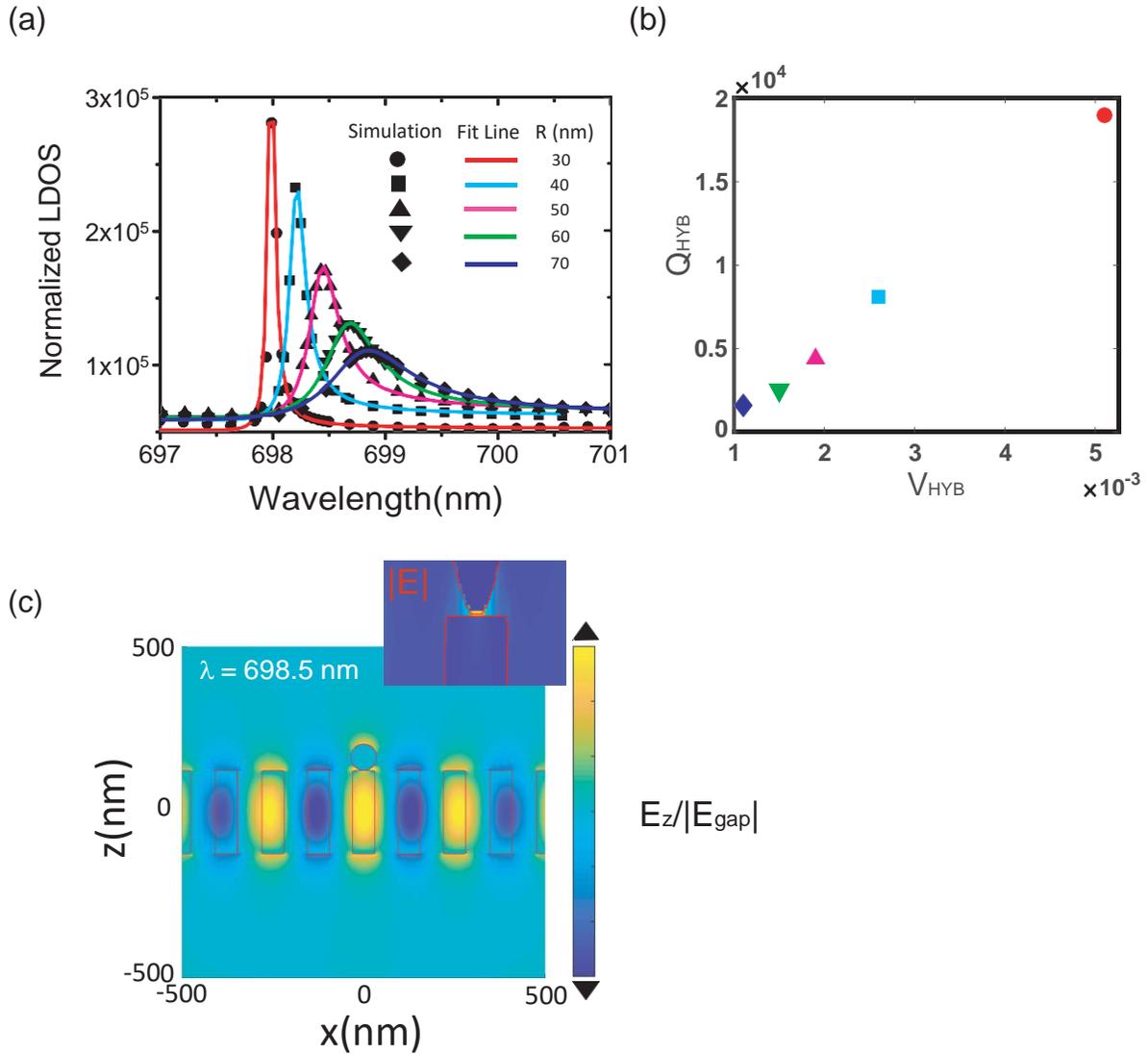

Figure 3: Simulation results of the hybrid cavity as a function of the gold nanosphere radius. (a) Normalized LDOS and (b) quality factor and normalized mode volume for the hybrid system constituted by the cavity beam and a gold nanoparticle of radius varying from $R = 30$ nm to 70 nm. The sphere is separated from the cavity by a gap of $d = 1$ nm. (c) Mode profile (ZX crosscut) of the $z$-component of the electric field normalized to the electric field amplitude in the gap ($|\mathbf{E}_{gap}|$ represented in the inset) for a $R = 40$ nm nanosphere-based hybrid cavity.

We also consider different shapes of the metallic nanoparticle to be coupled to the photonic crystal cavity: a nanocube and a nanoellipse (see Fig. 4). Metal nanocubes, which as well as nanospheres can be easily built by chemical procedures, are also able to play the role of the plasmonic particle in NPoM configurations.[31,41] Remarkably, the results in terms of $Q$ and $V$ are similar to the case of the nanosphere previously presented in Fig. 3. The nanoellipse, which also shows clearly the hybridization between the plasmonic and photonic resonances, is included to mimic the effect of a sharp metallic tip that may be placed near the SAM, as in other previous experiments for extreme light-matter interaction in sub-nm gaps.[42] This demonstrates that our proposed design is versatile and capable of functioning as a hybrid for different kinds of metallic nanoparticles.

To gain further insight into the hybridization of the metallic nanoparticle with the dielectric cavity, in Fig. 5, we plot the mode profile of the amplitude of the electric field ($|\mathbf{E}|$) for the XY crosscut (corresponding to the top surface of the cavity) at the position of the dipole (0.5 nm above the cavity). The analysis is performed for the three different studied geometries (sphere, ellipsoid and cube). For comparison, the field on the bare cavity is also included. Noticeably, when the metallic particle is considered, a subwavelength hot-spot is observed just below the nanoparticle. This hot-spot, which closely resembles the standard NPoM case, results from the squeezing of the evanescent vertical field in the photonic cavity by the plasmonic nanoparticle.

Besides the Purcell factor, it is also worthwhile to consider the radiative efficiency of the different hybrid modes. In the calculations above, we have considered both radiative as well as non radiative contributions to the LDOS. In order to ensure an efficient excitation to the environment (bright hybrid mode), it is desirable that the radiative contribution dominates over the non-radiative one. For the different analyzed geometries (sphere, ellipsoid, and cube), our simulations show that the ellipsoid and the spheres (specifically those with smaller radius) display a higher radiation efficiency (radiative vs. non-radiative ratio between 1 and 2) than the nanocubes. This behaviour can be explained attending to the hybridization of the TM fundamental mode of the cavity with the plasmonic mode excited in the NP. As the particle exhibits a more dipolar character, the hybridization between the modes of the cavity and the NP is larger, increasing the radiative power density with respect to the power loss density. Still, in the current scheme, the hybrid mode could be easily excited using TM guides modes of the dielectric waveguide, as usually done in direct inline coupling of photonic crystal cavities.

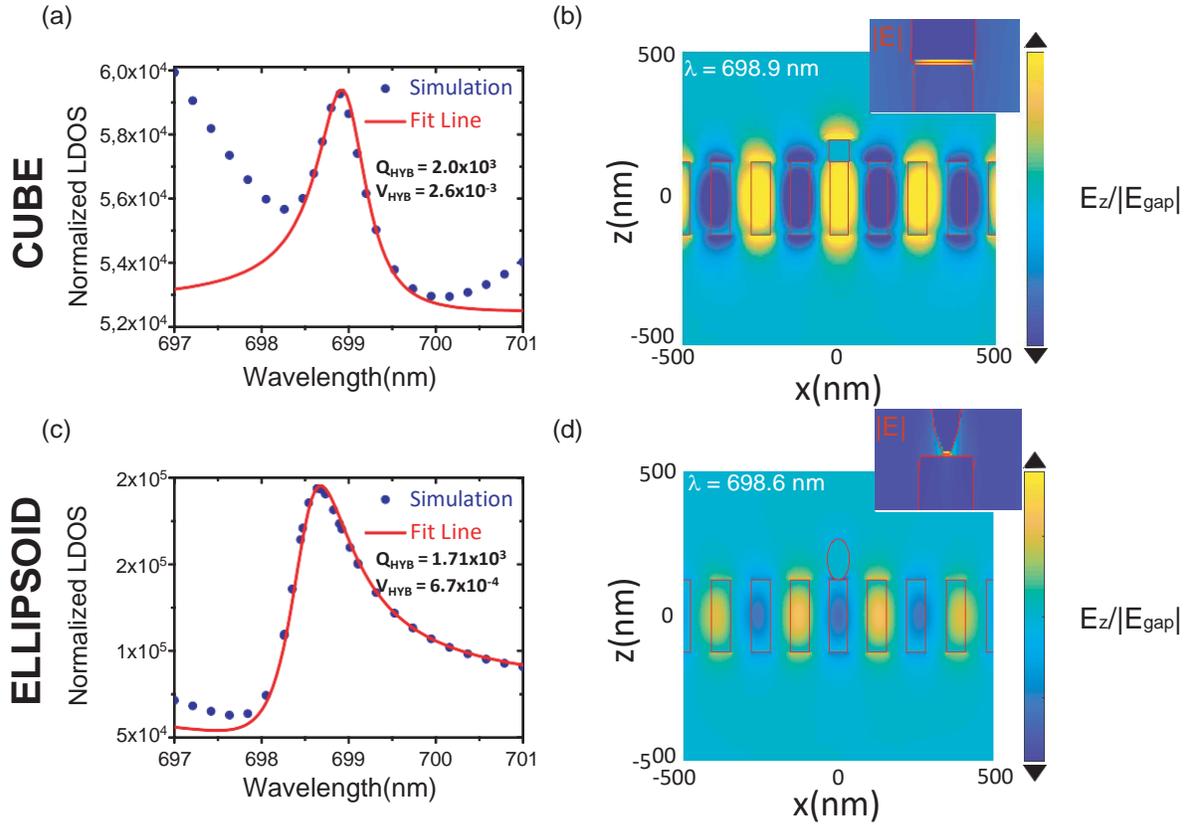

Figure 4: Simulation results of alternative configurations for the hybrid cavity. (a) Normalized LDOS, quality factor and normalized mode volume and (b) Mode profile (ZX crosscut) of the z-component of the electric field normalized to the electric field amplitude in the gap ($|\mathbf{E}_{gap,z}|$ represented in the inset) for a nanocube-based hybrid cavity. The length side of the cube is $l$ = 75 nm. (c,d) are as (a,b) but for a nanoellipse-based hybrid cavity. The width and length of the ellipsoid are $w_e$ = 40 nm and $l_e$ = 70 nm, respectively.

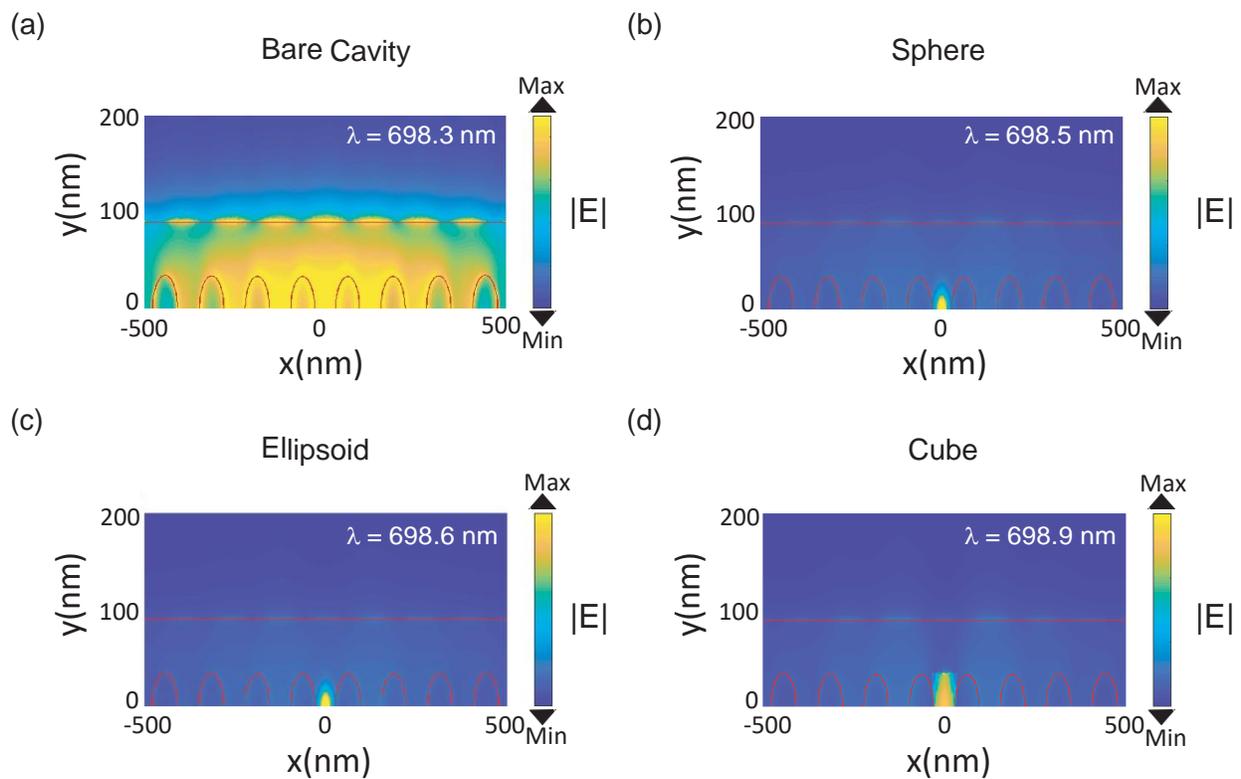

Figure 5: Mode profiles of the electric field amplitude |E| corresponding to the XY crosscut at the position of the dipole (0.5 nm above the cavity). The bare cavity (a), sphere (b), ellipsoid (c) and cube (d) cases are shown.

# Coupled harmonic oscillator (CHO) model

Recently, it has been shown in a theoretical work by Doeleman et al.[8] and confirmed in a numerical study in Ref. 11 that the behavior of hybrid plasmonic-photonic resonator systems can be accurately predicted by simple analytical modelling of antenna and cavity as a set of coupled harmonic oscillators, driven by a dipole source that represents an emitter. This coupled harmonic oscillator (CHO) model was verified to hold quantitatively for a large class of hybrid plasmonic-photonic resonators and predicts hybrid resonator performance by considering as input exclusively the properties of the bare antenna (scattering properties, LDOS enhancement in absence of the cavity) and cavity (mode volume, $Q$). In particular, the class of systems for which the model was verified includes photonic crystal nanobeams supporting TE confined modes coupled to plasmonic dipolar antennas (such as bow-tie antennas). The physics underlying the model is the assumption that antenna and cavity are weakly coupled, and therefore, the analysis of the antenna response can be properly described through its induced dipole moment. Here we verify whether the structures studied in this work, with their extremely small gaps, can be similarly treated.

To recapitulate the modelling approach, the total LDOS (LDOS$_{\text{tot}}$) for such hybrid systems can be expressed as[8]

$$\text{LDOS}_{\text{tot}} = 1 + \frac{6\pi\epsilon_0 c^3}{\omega^3 n} \text{Im}\left[\alpha_H G_{\text{bg}}^2 + 2G_{\text{bg}}\alpha_H\chi + \chi_H\right], \qquad (2)$$

where $n$ is the refractive index of the background medium, $c$ the speed of light and $\omega$ optical frequency. The LDOS$_{\text{tot}}$ is determined by the interference of three terms, each of which combine response functions of antenna and cavity, dressed by the multiple scattering interactions between them. The subscripted quantities $\alpha_H = \alpha/(1-\alpha\chi)$ and $\chi_H = \chi/(1-\alpha\chi)$ represent the hybrid polarizability of the antenna (antenna polarizability modified by the presence of the cavity), respectively the hybrid cavity response function as perturbed by the presence of the antenna. These hybrid quantities are in turn set by the bare antenna polarizability $\alpha$ which sets the antenna scattering and extinction cross sections in absence of the cavity structure, and the bare cavity response function $\chi$. Eq. 2 should match the LDOS provided by just the bare antenna if the cavity terms are removed (setting $\chi = 0$, also leading to $\alpha_H = \alpha$). Given that $\alpha$ is already fixed by the antenna scattering and extinction cross section, this matching of LDOS is accounted for by $G_{bg}$, which in a point dipole model is interpreted as the coupling between antenna and source as given by the Green's function in the surrounding background medium. In a similar spirit, in absence of the antenna $\alpha = 0$, the LDOS in Eq. 2 should match the bare cavity LDOS. This sets the Lorentzian response function of the bare cavity $\chi = \frac{1}{\epsilon_0 V_C}\omega^2/(\omega_C^2 - \omega^2 - i\omega\kappa)$ that has resonance frequency $\omega_C$ and damping rate $\kappa = \omega_C/Q$ matching the resonance and $Q$ of the unperturbed cavity, and in which the effective mode volume $V_C$ appears such that Eq. 2 matches the Purcell factor of the cavity at the assumed location of the emitter.

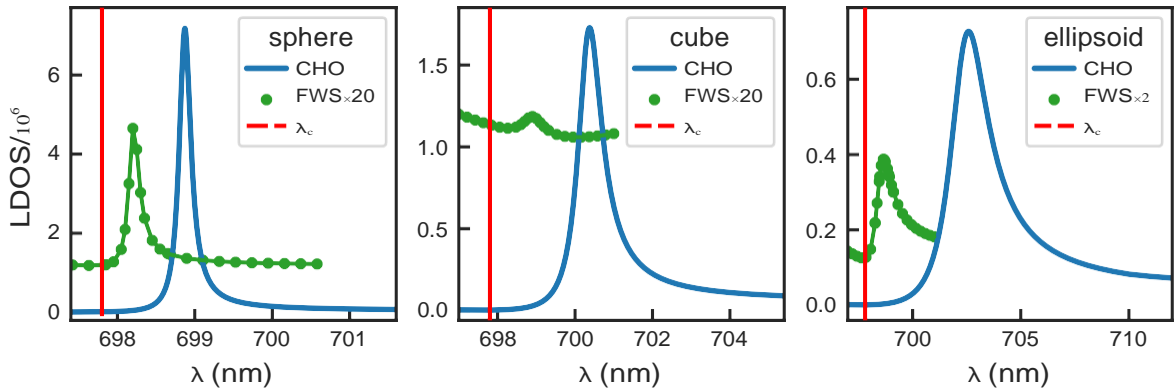

Figure 6: Comparison of the LDOS as a function of the wavelength for the three hybrid systems considered. The green line shows the results of the full-wave simulations (labelled FWS) discussed above. The blue curve shows the LDOS as calculated with Eq. 2 (labelled CHO for coupled oscillator model). The resonance frequency of the cavity in absence of the nanoparticle is indicated as $\lambda_c$ (dashed vertical lines). The full-wave simulations have been scaled as indicated.

In order to investigate the agreement between our COMSOL simulations and the CHO model, we performed additional simulations on the antennas discussed above, again using COMSOL Multiphysics, to calculate the scattering and extinction cross sections of each antenna in absence of the cavity and by using plane wave excitation. Within the assumption of dipolar scattering, the obtained frequency-dependent scattering and extinction cross sections can be used to obtain the antenna polarizability by inversion of the relations[43]

$$\sigma_{\text{scat}} = \frac{\omega^4}{6\pi\epsilon^2 c^4}|\alpha|^2, \quad \sigma_{\text{ext}} = \frac{\omega}{\epsilon^0 cn}\text{Im}(\alpha). \quad (3)$$

Next, once we have determined $\alpha$ from the far-field scattering properties we determine $G_{\text{bg}}$ (which is approximately real-valued in the near field) from COMSOL calculations of the total LDOS provided at the emitter by inverting Eq. 2 (setting the cavity terms to 0). For a benchmarking of this procedure for dipolar gap-antennas we refer the reader to Refs. 8, 11. We note that the inversion of scattering properties into polarizability only works in frequency domains where the cross sections are in magnitude within the constraints of the unitary limit that provides an upper bound on extinction of dipolar scatterers ($\sigma_{\text{ext}} \leq 3/2\pi\lambda^2$). While some of the antennas at hand do not satisfy this constraint at their resonance frequency, the antenna-cavity hybrids operate significantly to the red of the antenna resonance frequencies. At the relevant operation frequencies all the antennas show far-field scattering properties well described as electric dipole scattering.

In Figure 6 we show the LDOS calculated from the CHO model, evaluating Eq. 2 with parameters for the bare constituents extracted from the COMSOL simulations. We find that the CHO model (blue lines in Figure 6) predicts Fano lineshapes in LDOS for all the hybrids at hand that are qualitatively similar to those calculated with full-wave simulations of LDOS for the full hybrid system (green curves/symbols). Quantitatively, the CHO prediction significantly overestimates the maximum achievable LDOS of the hybrid systems by a factor of 10—50. In addition, the full-wave numerical simulations are shifted significantly less than the CHO model suggests in resonance wavelength, feature a different degree of broadening, and generally a poorer contrast between peak LDOS and background LDOS (LDOS a few linewidths away from the Fano resonance). This should be contrasted to earlier findings by Palstra et al.[11] wherein the very same procedure gave an excellent quantitative match between the CHO semi-analytical model and full-wave simulations of hybrid LDOS, in the case of TE-cavity modes in photonic nanobeams coupled to dipolar gap antennas.

The discrepancy between the CHO model and the full-wave results points to the fact that the NPoM-cavity hybrids operate in a qualitatively different regime than previously reported hybrid plasmonic-photonic resonators constructed from dipole antennas and photonic crystal nanobeams. At the extremely small gaps considered in this work, the antennas do not respond with only their electric dipole mode as assumed in the CHO model, and instead high multipole orders in their excitation also contribute. Even without examining the hybrid LDOS spectra, the failure of the dipole model is already evident from the anomalously small frequency shift induced in the cavity mode by the antennas. In cavity perturbation theory a seminal result due originally to Waldron and Bethe is that the complex resonance frequency of the unperturbed cavity $\tilde{\omega}_c = \omega_0 + i\kappa/2$, where as before $\kappa$ is the FWHM of the unperturbed cavity resonance, will shift by $\Delta\tilde{\omega}$ following

$$\frac{\Delta\tilde{\omega}}{\tilde{\omega}} = -\frac{\alpha}{2V_C}, \qquad (4)$$

where $V_C$ is the mode volume of the unperturbed cavity.[44–48] In more recent literature it was shown that this expression generalizes to a wide class of photonic resonances, provided the mode volume $V_C$ is generalized through quasi-normal mode concepts.[49–51] In this perturbation formula the real part of antenna polarizability $\alpha$ results in a frequency shift, and the imaginary part in additional resonance broadening. It can be analytically verified that the CHO model intrinsically reproduces exactly this well-known expression. In other words, the CHO model predicts the hybrid LDOS Fano resonance to occur at the perturbed cavity resonance predicted by Eq. 4. Comparing the resonance shifts between COMSOL simulation and CHO model in Figure 6 thus directly shows that in all the cavity-antenna constructs the frequency shifts are far smaller than would be expected on basis of perturbation theory and the antenna far-field scattering properties that are captured by the polarizability *α*. These discrepancies can be directly ascribed to the fact that Waldron's formula assumes the perturbation to be placed in a part of the cavity field where the gradient is small, so that the perturber provides an essentially quasistatic dipolar response. Instead, the field plots directly show very strong field gradients in/at the antenna. Such gradients, and particularly the extreme gap fields are only supported by virtue of retarded and multipolar contributions.[51] We thus conclude that the NPoM inspired hybrids proposed in this work are qualitatively distinct from the previously reported microdisk and nanobeam resonator hybrids due to significant multipolar corrections.

## Discussion and conclusion

It is useful to provide a perspective on the achievable *Q* and *V* values in the proposed NPoM-inspired hybrids. Figure 7 depicts the *Q* vs. *V* values for the different structures under study in this work. Through this diagram it is possible to compare the hybrid systems and the isolated constituents, in correspondence with Ref. 11. As expected, the hybrid systems are in the intermediate regime, taking *Q* and *V* values between those of the photonic cavity and the plasmonic nanoparticle. Dashed diagonal lines show lines of constant Purcell factor. At the very small emitter-particle distances considered in this work, the LDOS values and *Q*'s for the bare nanoparticles and for the sphere on a non-structured substrate are quite similar. Our hybrid cavities enable values of $F_P$ above $10^5$. For these extremely narrow gaps the LDOS-enhancement in terms of achievable peak LDOS are less advantageous than extrapolated from the CHO model. At the same time the achievable LDOS values are higher than those that have been predicted for hybrids composed of standard TE-polarized photonic crystal nanobeam cavities and bow-tie antennas,[11] for bow-tie gaps that are realistically achievable by lithography (limited to circa 15 nm). To attain LDOS values in such TE hybrids based on bow-tie nanoantennas that are as high as we find in this work for NPoM-inspired hybrids, would require bow-tie gaps below 5 nm, which are challenging to achieve in a controllable fashion, as discussed above.

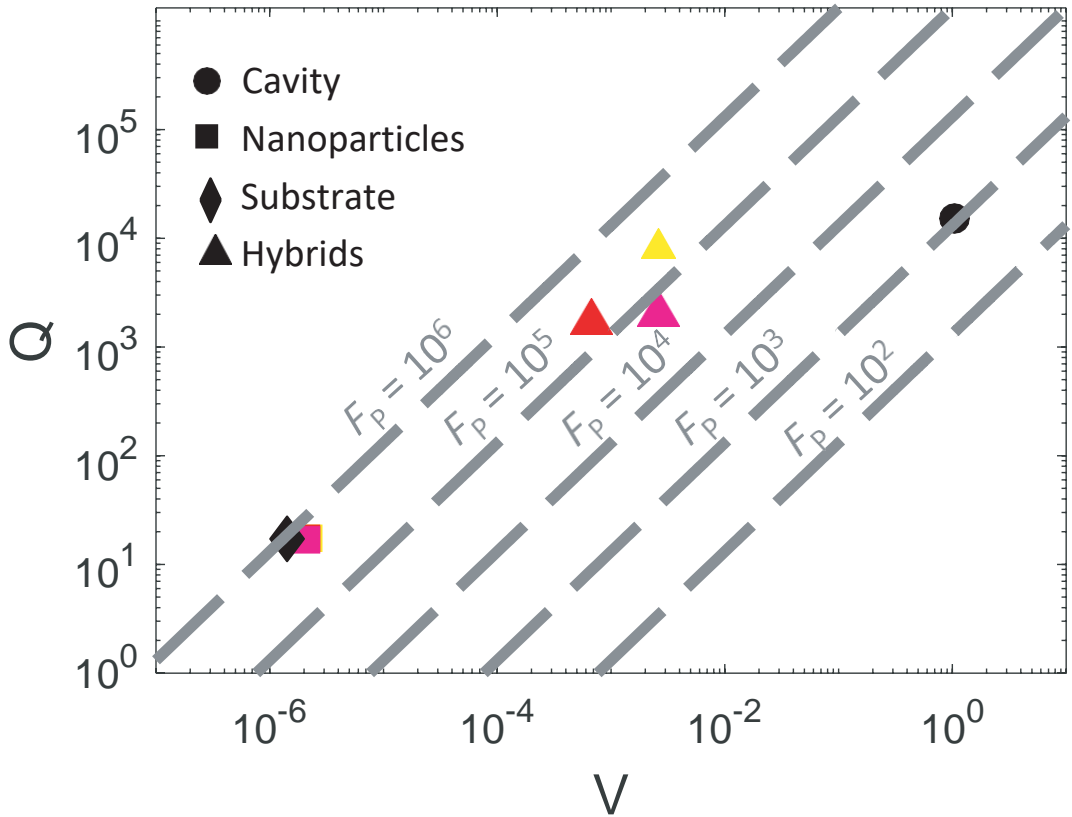

Figure 7: Quality factors $Q$ and normalized effective mode volumes $V$ for bare metallic nanoparticles (squares of different colors according to the nanoparticle geometry: sphere–yellow; ellipsoid–red and cube–magenta), the photonic crystal nanobeam cavity (circle), the spherical nanoparticle ($R$ = 40 nm) on a non-structured gallium phosphide substrate (diamond) and the hybrids (triangles of different colors according to the geometry of the nanoparticle on the cavity: sphere–yellow; ellipsoid–red and cube–magenta). The dimensions for the different nanoparticles are: sphere ($R$ = 40 nm), ellipsoid ($w_e$ = 40 nm and $l_e$ = 70 nm) and cube ($l$ = 75 nm). Diagonal dashed lines are lines of constant Purcell factor with value $F_p$ as labelled.

In summary, we propose a novel design of hybrid photonic-plasmonic cavities based on the combination of a NPoM plasmonic cavity and a TM dielectric photonic crystal cavity. This nanostructure is suggested as an alternative to previous hybrid cavities, which are realized by means of metallic bow-tie antennas on dielectric cavities. Our proposed hybrid cavity works for different geometries of metallic nanoparticles including spheres, ellipsoids or cubes. The main advantage of the proposed hybrid is that it enables nm- and sub-nm scale gaps in a controllable way, which is out of reach in a repetitive way when the gap is defined lithographically. Such nm-scale gaps are required to get extreme small mode volumes, eventually reaching the picoscale as in standard NPoM approaches. We evidence that for a gold nanoparticle separated from a dielectric photonic cavity by a 1 nm gap, normalized mode volumes around $10^{-3}$ are achieved. Furthermore, the $Q$-factor exhibit high values (around $10^3$), which could even be larger by implementation of a more exhaustive design of the photonic cavity. It is the first time, to the best of our knowledge, that it is proposed a feasible hybrid plasmonic-photonic cavity with high Purcell factors ($F_P = 10^5$). In comparison with the Purcell factors obtained with feasible gaps ($d = 10$ nm) for hybrid plasmonic-photonic (TE) configurations, we reach an improvement of one order of magnitude. This means that this hybrid configuration is able to combine the best of both configurations: high $Q$-factors due to the dielectric cavity and extremely small mode volumes (well below the diffraction limit) thanks to the metallic nanoparticle. In addition, as it was demonstrated in previous works based on bow-ties antennas on dielectric cavities operating for TE polarization, changing the detuning between the cavity and the nanoparticle it is possible to achieve different $Q$ and $V$ values. In our approach, by increasing the radius of the nanosphere from $R = 30$ nm to $R = 70$ nm, $Q$ and $V$ range from $Q_{HYB} = 1.9 \times 10^4$; $V_{HYB} = 5.1 \times 10^{-3}$ to $Q_{HYB} = 1.6 \times 10^3$; $V_{HYB} = 1.1 \times 10^{-3}$. Changing some parameters of the photonic cavity, such as the period or the nanobeam thickness, should also enable a fine tuning of $Q$ and $V$. Our results unveil a new building block in the context of hybrid plasmonic-photonic circuits[52] which should find application in enhanced Raman scattering, harmonic generation and molecular optomechanics even in the few-photon regime.

## Methods

### Numerical simulations

The results are calculated by means of the Finite Element Method (FEM), implemented in the commercial software COMSOL Multiphysics.[53] Using the Radio Frequency Module in COMSOL, the Maxwell equations together with the boundary conditions are solved in the frequency domain. The optical constants for gold nanoparticles are taken from Ref. 54. The refractive index for gallium phosphide (GaP) is $n = 3.49$. To determine the LDOS, the structure is illuminated by an electric point-dipole source, whose dipole moment is considered along the z-axis (along the axis connecting the sphere and the cavity). The normalized local density of optical states is defined as:

$$\text{LDOS} = \frac{P_{rad} + P_{loss}}{P_{rad\ air} + P_{loss\ air}} \quad (5)$$

where $P_{rad}$ is the radiative power density and $P_{loss}$ is the total power loss density. The numerator corresponds to the radiative and nonradiative power emitted by the dipole coupled with the structure. However, the denominator contains the radiative and nonradiative power emitted by the dipole in vacuum. The dipole is located at the center of the structure and in the middle of the gap between the nanoparticle and the cavity. This means 0.5 nm below the nanoparticle and 0.5 nm above the surface of the beam. The hybrid geometry is surrounded by a cylindrical air region of radius 1.72 μm. An additional smaller cylinder with a radius of 1.11 μm and made of air is placed in the center of the larger cylinder. The scattered power ($P_{rad}$) is calculated at the boundaries of the smaller cylinder, whereas the total power loss density is obtained by means of the volume integration of the losses in the metallic nanoparticle. The dipole source is surrounded by a sphere with a diameter which is equal to the gap size ($d$ = 1 nm) to ensure a sufficient fine grid in close proximity to the dipole source. The mesh of the surrounding air medium is chosen to be smaller than 150 nm and that of the cavity and particle is smaller than 45 nm. The mesh of the sphere surrounding the dipole source is smaller than 1.2 nm. The cylindrical region of air is surrounded by a perfectly matched layer (PML) with a thickness of 0.5 µm. The mode profiles are calculated with the eigenmode solver of COMSOL Multiphysics. Through these near-field plots we can see the electric field distribution in the structure for the eigenfrequencies of interest. It is remarkable to clarify that the same mesh is used for the LDOS and eigenmode calculations and for the different analyzed structures: bare nanoparticles, bare cavity and hybrid. In all the cases, the same geometry is considered and only optical constants of the materials are changed appropriately. For the near-field map corresponding to the isolated sphere, the structure is illuminated with a plane wave propagating along the *x*-axis and linearly polarized along the *z*-axis. As the main aim is to observe the kind of plasmonic resonance that is excited (in this case dipolar electric), plane wave illumination is chosen. Furthermore, attending to time-consuming computations, the nanosphere is located in a homogeneous spherical air region of radius 700 nm. The PML thickness corresponds to 200 nm. To ensure numerical convergence of the results, the tetrahedral mesh is chosen to be sufficiently fine. Thus, the mesh of the nanosphere is smaller than 10 nm.

## Acknowledgement

We thank C. Galland for useful discussions. This research has been funded by the European Commission under the Project THOR (H2020-EU-829067). A.B. acknowledges financial support by the Alexander von Humboldt Foundation. A. M. acknowledges support from the Spanish Ministerio de Ciencia, Innovación y Universidades (grants PGC2018-094490-B, PRX18/00126) and the Generalitat Valenciana (grants PROMETEO/2019/123, PPC/2018/002).

**Graphical TOC Entry**

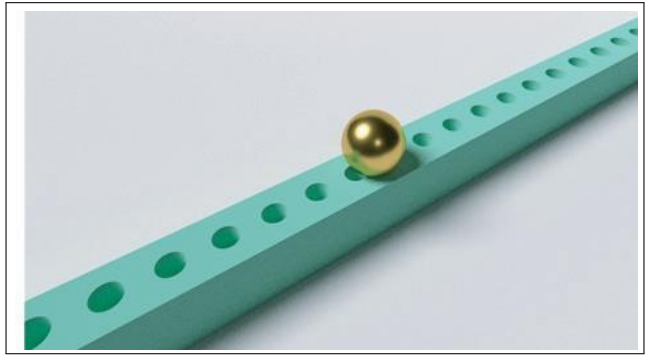